%Paper: hep-ph/9404205
%From: rmohapatra@umdhep.umd.edu
%Date: Fri, 01 Apr 1994 11:38:05 EST

\font\titlefont = cmr10 scaled \magstep2
\magnification=\magstep1
\vsize=20truecm
\voffset=1.75truecm
\hsize=14truecm
\hoffset=1.75truecm
\baselineskip=20pt

\settabs 18 \columns

\def\b{\bigskip}
\def\bb{\bigskip\bigskip}

\def\ce{\centerline}

\def\no{\noindent}

%$$\eqalign{
% put in lines of equations here, each ending in \cr
%}$$

%$$\eqalign{
%put in equations here ending each line with \cr
%} \eqno (1)$$
%the above will put the one between the two lines of equations and set it
%off to the right

% END BEGINNING FORMATS % BEGIN HEADER
\rightline{ UMD-PP-94-116}
\rightline{ March, 1994}
\ce{\titlefont{ Understanding the Standard Model }}

\bb

\ce{\bf{ R.N. Mohapatra\footnote\dag{\rm{ Dedicated to the memory
of Robert E. Marshak, to be published in the memorial volume
by George Sudarshan.  }},~~   }}

\ce{\it{ Department of Physics, University of Maryland,
College Park, MD 20742}}

\b
\ce{\bf Abstract}

\no

   Freedom from Adler-Bell-Jackiw anomalies is a primary requirement
for the renormalizability of a gauge theory of chiral fermions, which
forms the basis of the successful standard model of electroweak
interactions and its many extensions. In this article, we explore
to what extent, the assumptions behind the standard model
as well as the observed quantization of electric charges
of quarks and leptons
can be understood using the various anomaly constraints and
how the situation changes as one tries to incorporate a
 nonvanishing neutrino mass .

 \filbreak

{\bf {I. Introduction:}}

          Robert E. Marshak will be remembered for his many
contributions to our understanding of particle physics such as
the V-A theory of weak interactions with George Sudarshan, the
two meson hypothesis with Hans Bethe etc. These ideas are
part of the history of particle physics and will be remembered
far beyond the time when our generation of physicists are gone.
Those of us, who were lucky enough to have been associated with
him and to have shared  his enthusiasm for physics and life, will
always remember him not only for his deep intuition into physics but
also for his generosity of spirit, for being a constant voice of reason
and for his unflagging commitment to better the world in every way he could.

My association with Bob Marshak dates back some
 twentyfive years, starting first as a student at Rochester, then as
a colleague and collaborator at CCNY and Virginia Polytechnic Institute
at Blacksburg. During this period, we bounced many ideas off each other,
discarded most but  advanced a few in print.
I want to take this opportunity to briefly recall one  idea
 that we particularly felt very strongly about.
 It has to do with the suggestion
 that physics beyond the standard model must contain
local $B-L$ as a symmetry of nature[1]. This idea originated
 in 1979 and 1980
[1] from the strong prejudice that the standard model must be
extended to include the right-handed neutrino to make it completely
quark-lepton symmetric. Marshak was of course one of the inventors,
along with Gamba and Okubo,
of the idea of Baryon-Lepton symmetry [2] which in the modern language
becomes quark-Lepton symmetry and the idea had already been successfully
used by Bjorken and Glashow[3] to predict the existence of the charm quark.
Extending the same quark lepton symmetry
to the standard model leads naturally to a left-right symmetric ( LRS )
model of weak interaction , where the $U(1)$
symmetry is nothing but the $B-L$ generator.( Although
the LRS model had already been discussed by this author
in collaboration with Pati, Salam and Senjanovi\'c[4], the identification
of the $U(1)$ generator with $B-L$ was not known ).
In the standard model, due to the
 absence of the right-handed neutrino, the
$B-L$ is an anomaly-free global symmetry but since $Tr[B-L]^3\neq~0$,
this symmetry cannot be gauged; however, in the presence of the $\nu_R$,
the cubic $B-L$ anomaly vanishes making $B-L$
a gaugeable symmetry. Now, if we "subtract" $B-L$ from the weak hypercharge,
the "left-over" piece can be identified with the right-handed weak
isospin $I_{3R}$ and it then becomes possible to
 recast the electric charge formula in a more physical form[1]

$$Q = I_{3L}~+~I_{3R}~+~{{B-L}\over {2}}\;\eqno(1)$$

Once $I_{3R}$ emerges as a gauge symmetry, it is natural to
gauge the entire right-handed gauge symmetry present among the
fermions and of course this leads to the LRS models.
The formula in equation 1 has several
 interestng consequences such as
A) the neutrino being a Majorana particle with the smallness of
 its mass being
connected to the scale of spontaneous parity violation[5],
B) the possible existence of a new baryon violating
process , the neutron-anti-neutron oscillation[6] etc. which
under certain circumstances may be observable.
Heroic experiments were carried out at Grenoble[7] to search for
 the neutron-anti-neutron oscillation; no evidence for this process
were found at the level of their precision; however it
yielded the best upper limit on the process.
On the other hand, the idea of a
Majorana neutrino is now-a-days an integral part of the discussion
in grand unified theories as well as neutrino phenomenology.
 The experimental results in this  front
may be more hopeful.
If recent indications of possible non-vanishing neutrino mass
in several experiments such as the solar neutrino observations
as well as the atmospheric neutrino data is confirmed in the future,
 a spontaneously broken local $B-L$ symmetry as an integral part of
physics beyond the standard model will be confirmed.
Further possibilities for unification that can
incorporate the $B-L$ symmetry, such as  SO(10) grandunification,
, and preon models etc, were discussed during the past decade
and have been summarized in the
excellent book[8] by Marshak,
 finished only a few days prior to his death. I
will not discuss these ideas any further in this article. Instead
I want to focus on another of Bob Marshak's obsessions during the
last five years of his life, having to do with a better understanding
of the origin of the immensely successful standard model[9].

In late eighties, Marshak along with C. Q. Geng[10] began a program
 of trying to understand the assumptions
behind the standard model starting from
the requirement of anomaly cancellation[11].
Their work was subsequently followed
up by a number of authors including this one[ 12, 13, 14, 15 ].
 These works have led to several
remarkable results that throw light on ( and even clarify ) some of the
basic assumptions on which the standard model rests. In this article,
I will discuss the present status of this approach.

\bb

{\bf {II. Statement of the problem:}}

In discussing the standard model[9], one usually postulates the following:

a) the
known set of fermions $(u^{\alpha}, d^{\alpha}, e, \nu )$ for each generation
( where the superscript $\alpha = 1, 2, 3$ denotes the color index );

b) the local gauge symmetry of nature is
 $SU(3)_c \times SU(2)_L\times U(1)_Y$ ;

c) and the gauge quantum numbers for the fermions are
so chosen as to match their observed properties e.g. the left-handed
up and down quarks are part of an $SU(2)_L$ doublet ( as are the
$(\nu_L,~e_L)$ ) to produce the correct beta decay interactions; their
$Y$ quantum numbers are so chosen as to reproduce the correct values
of the electric charges etc.

In view of the extraordinary success of this model, it is interesting to
ask whether it is possible to reproduce these assumptions
 starting from a more economical principle.

 One may consider two approaches to this problem; in the first
 one to be called the top-down approach, one may start
from some general principles ( perhaps String theory , Higher dimensional
theory etc ), and derive a theory of forces and matter at a high scale
such as the Planck scale and as this theory is reduced to represent
physics at lower energies, the standard model may emerge. The heterotic
string theory of Gross, Harvey, Martinec and Rohm[16] could be such
a theory- but as is well known, it does not lead to a unique theory
at low energies . If eventually it does, then it would provide an
immensely satisfactory basis for the standard model and practically,
pre-empt any other approach. A less ambitious point of view in the same
top-down philosophy, is to start with a grandunified theory; but as
is well-known, the nature of the low energy physics in the grand unified
theories depends a great deal on the choice of Higgs multiplets that are
used to to break the model down to the standard model. We will briefly
remark on this in a subsequent section.

 Another approach, which could be called
the "bottom-up" approach, is to proceed from the low energy side, using
again some general principle and see how much of the assumptions behind
the standard model  can be understood. It is this approach, that we will
discuss in this article. It is perhaps fair to point out that the top-down
approach has one advantage over the bottom-up philosophy i.e.
 the former  can generate  detailed dynamics whereas the latter has
so far only been used to
shed light on the "geometrical" properties  such as quantum numbers and
gauge groups etc.

\bb

{\bf III. Standard model quantum numbers from Anomalies:}

Our basic tool will be the freedom from Adler-Bell-Jackiw anomalies[13]
required for the consistent renormalization and unitarity of a gauge
field theory. I will not follow the historical path ; rather, will
will follow the discussion of a recent paper by Frampton and I[17]
and gradually embed the earlier discussions where they fit in.
It is clear that anomalies alone cannot lead to a definite theory
without some way to specify the underlying chiral fermions and/or
some knowledge of the gauge symmetry that is responsible
for the dynamics. Since we are exploring the nature of weak
interactions, it may not be too objectionable  to assume the gauge
symmetry of strong interaction as a starting point.
Therefore, we take as  our starting
assumptions the following:

i) the local symmetry of electroweak and strong
interactions at high energies is given by $SU(3)_c\times U(1)_Y\times
G^{\prime}$;

ii)  $SU(3)_c$ is vector-like;

iii) and that the set of Y-charges of the fermions is irreducible.

 The first two
assumptions are self-explanatory but the third needs to be explained.
By {\it irreducible} set of Y-charges, we mean that if any one or two
Y-charges are such that a gauge invariant mass is allowed for the fermion
or the fermion pair, then we forbid such an assignment. The physical argument
behind this assumption is that the value of the gauge invariant mass
 can be arbitrarily large compared to the scale of the gauge symmetry
( i.e. $U(1)_Y \times G^{\prime}$ ) and the fermions then will not be
part of the low energy spectrum.

We will show that the above assumptions
combined with the requirement that the gauge group be anomaly-free
leads to the following conclusions[17]:

(a) The minimal number of fermions that leads to an
anomaly-free theory is 15, which is precisely the
number of fermions in the one-generation standard model;

(b) The maximal allowed simple $G^{\prime}$ is $SU(2)$ which must be
parity-violating;

This set of assumptions  appears more economical
than those made in the usual construction of the standard model and the
fact that one can reproduce two of the key ingredients of  the one generation
standard model i.e. the number of fermions, the weak gauge group
is quite intriguing.
Supplemented by an extra assumption that QED is vector-like,
enables one reproduce the quantum numbers of the basic fermions of the model
, thereby , explaining one of the major mysteries of theoretical
physics i.e. why are the observed electric charges quantized ?

In order to prove the above assertion, let us say that
 the number of fermions is N
and is divided into two groups called quarks and leptons, the quarks
being defined as triplets or anti-triplets under color and leptons being
singlets. The assumed vector nature of QCD requires the number of triplets and
antitriplets to be the same; let this number be equal to $Q/2$ where Q is
hence an even number. Denoting the number of leptons by $L$, one has $N=3Q+L$.
Let the leptons and quarks have Y-charges $y_i$ $(1\leq i \leq L)$ and
$z_j$ $(1 \leq j \leq Q)$ respectively. The three anomaly constraints
arising from  $U(1)[Gravity]^2$, $U(1)[SU(3)_c]^2$ and $[U(1)_Y]^3$
lead to the following equations:
$$\Sigma_1^L ~y_i = 0;\eqno(2a)$$

$$\Sigma^Q_1 ~z_j = 0;\eqno(2b)$$

$$\Sigma^L_1 ~y^3_i + 3~ \Sigma^Q_1~ z^3_j = 0;\eqno(3)$$

 We will be interested in finding the smallest N for which the
$y_i$ and $z_j$ will satisfy Eqs.(1)-(3).  The assumption of
irreducibility implies that $L\geq3$.
Further since Q is even it can be $2, 4,
6$ etc.
  If $Q = 2$, by Eq. (2b) $z_1~=~-z_2 $ so that the two quarks
are allowed to
acquire gauge invariant mass. ( One might wonder about invariance
under $G^{\prime}$; but
 the only acceptable $G^{\prime}$
in this case is $U(1)$ and anomaly freedom implies that its charges
must also be equal and opposite.) Thus we expect this quark pair then
to decouple from the low energy spectrum. So, the
 smallest value of Q is  4 leading to
$N = 15$. This proves the first assertion above.
In the rest of the paper, we will denote the quark hypercharges by
$(z_1,~z_2,~z_3,~z_4)$ and the leptonic ones by $(y_1,~y_2,~y_3)$.

Let us now show that the maximal group $G^{\prime}$ is $SU(2)$.
First we show that $G^{\prime}\neq SU(3)$.
Since this $SU(3)$ must be orthogonal to color $SU(3)_c$, the three
leptons must be a triplet under it in which case, Eq.(2a)
implies that $y_i = 0$ and that there is no way to satisfy the
anomalies arising from  the $G^{\prime}$ group.
  This leaves as the only possible simple group $G^{\prime}$=$SU(2)$.

As a brief digression, suppose $G^{\prime}$ were not simple but merely
a $U(1)$. It was  shown in ref.17 that
 the extra anomaly constraints associated
with it imply that it is vectorlike . To see this, take the
linear combinations of the two $U(1)$ charges to define two new $U(1)$'s
and call their charges X and Y respectively. By means of an appropriate
choice we can make it vanish for one of the leptons. Let us denote
the X-charges of quarks to be $x_a$ where $a = 1~~to~~4$ and those of
leptons to be $x_5, ~x_6~,0$ . The mixed $U(1)[Gravity]$ and
$U(1)[SU(3)]^2$ anomalies then imply :

$$x_1 + x_2 + x_3 + x_4 = 0;\eqno(4a)$$

$$x_5 =- x_6;\eqno(4b)$$

$$x^3_1+x^3_2+x^3_3+x^3_4=0;\eqno(4c)$$

Again as before if we choose the X charges also to be rational,
then Eq.(4), combined with the Fermat's last theorem will imply that
$x_1=-x_2$ and $x_3=-x_4$ i.e. $U(1)_X$ is parity conserving.

The next question then arises : how do the quarks and leptons transform
under this $SU(2)$ group ? Consistent with our assumptions,
there can be at most one doublet among the quarks otherwise Eq.(2) will
imply that the Y-charges for the quarks become reducible and the
quarks will acquire an invariant mass and decouple from the low
energy spectrum. The $G^{\prime}\equiv~SU(2)$ anomaly freedom then
implies that there has to be one lepton doublet.
  There are now five Y-charges
which describe the quark-lepton system ( using $z_1~=~z_2$ and
$y_1~=~y_2$  since they are members of the $SU(2)$ doublet ).
 Now  we[17] further demand that
electromagnetism is vector-like subsequent to spontaneous symmetry
breaking . We will show below that this  together with eq.3 leads to a
complete determination of the Y-charges and electric charge quantization.
To establish this result, first note that the generator of the
final unbroken $U(1)$ can be written in general as:

$$Q_e = I_{3L} + \eta Y;\eqno(5)$$

The constraints of vector-like $Q_e$ are:

$$-{{1}\over{2}}+\eta y_1 = 2\eta y_1;\eqno(6a)$$

$$+{{1}\over{2}}+\eta z_1 = -\eta z_3;\eqno(6b)$$

$$-{{1}\over{2}} +\eta z_1 = +\eta (2z_1 + z_3);\eqno(6c)$$

Equation (6b) and (6c) are actually equivalent; taken together,
eqs.(6) imply that $\eta = -1/{2 y_1}$ and $ y_1 = z_1 + z_3 $.
Putting these relations in eq.(3), we get $ y_1 = -3 z_1$ and
$ z_3 = -4 z_1$ as one solution that we use below. The other
solutions of the cubic anomaly equation (3) are eliminated by
the requirement of irreducibility.
If we now rewrite Y as $Y^{\prime} = 2 \eta Y$,
and call $Y^{\prime} $ as the standard model Y, then the electric
charge formula becomes

$$ Q_e = I_{3L} + {{Y}\over{2}}.\eqno(7) $$

The  values of the new Y are precisely those of the standard model.
Note also that in deriving the above electric charge formula nowhere
have we used any assumption about the Higgs structure of the theory
nor have we made use of the freedom from $SU(2)$ anomalies in this
derivation. In this sense, this is slightly different from the
discussion in ref.13.

\bb

{\bf {IV. Effect of non-zero neutrino mass:}}

As is well known, the standard model leads to vanishing mass for the
neutrinos due to the absence of the right-handed neutrino in the
theory. There are several reasons to include the right-handed neutrino
into the theory: one is the aesthetic reason of restoring quark-lepton
symmetry to weak interactions as discussed in the introduction; a
second reason ( of course much more compelling ) is the accumulation
of many indications for the existence of a non-zero neutrino mass
such as solar neutrino puzzle, a hot component to the dark matter
profile of the universe etc. The simplest way to understand the
neutrino mass is to include the right-handed neutrino. In this section,
we will discuss the impact of including the right-handed neutrino
on the anomaly discussion of the previous section.

In the presence of $\nu_R$, the minimal number of chiral fermions becomes
sixteen and the Y-set becomes $( z_1,~z_2,~z_3,~z_4)$ in the quark
sector and $(y_1,~y_2,~y_3~y_4)$ in the lepton sector. Let us first
discuss the question of possible choice for the gauge group $G^{\prime}$.
The anomaly  and irreducibility ( or decoupling ) requirements clearly
rule out both $SU(4)$ and $SU(3)$ as possible candidates. The only
allowed ones are $SU(2)$ and $SU(2)\times SU(2)$. Let us focus on
the case of $SU(2)$[13,14]; the case of $SU(2)\times SU(2)$ is
straightforward[14], the only modification being that to ensure
vector-like electric charges one must use the appropriate charge
formula that reads $Q~=~I_{3L}+I_{3R}+\eta Y$ instead of eq.5.

Note that the constraint equations in this case are
the same as in eq. 2 and 3
except that the equation involving y's includes the $y_4$ in the
sum over the $y$'s. Due to the
$SU(2)$ assignment, we have as before $y_1~=y_2$ and $z_1~=~z_2$. Using the
electric charge formula in eq. 5 and using the vector-like condition for
electric charge and the cubic anomaly equation,
 one finds the following solutions for
the hypercharges:
$ y_1~=~-3z_1;~y_3~=~{{1}\over{2\eta}}-y_1;~y_4~=~-{{1}\over{2\eta}}-y_1;
z_3~=~{{1}\over{2\eta}}-z_1; z_4~=~-{{1}\over{2\eta}}-z_1$.
All anomaly constraints and constraints from the requirement
of vector-like QED are automatically satisfied by the above
relations and there is no
way to rescale the hypercharge coupling to fix the hypercharge
quantum numbers uniquely as we did in the case of the standard
model. The electric charge quantization, therefore, does not follow
 in this case. The reason is traceable to the fact that now the
$B-L$ exists as a hidden gaugeable symmetry[14, 18] in this model.
 It was suggested
by Babu and this author[14] that the way to restore charge quantization
is to get rid of this hidden symmetry which amounts to having neutrino
as a Majorana particle. This can be explicitly seen from the
above relation involving the $y$'s and $z$'s as follows.
The only gauge invariant Majorana mass possible is for the $\nu_R$ and that
requires that $z_4~=~0$; this enables  all the $Y$ quantum
numbers to be expressible in terms of $1\over{\eta}$ and as in the
previous section redefining $Y$ in the same way leads to determination
of all $Y$ quantum numbers as well as to electric charge quantization.

{\bf {V.Grand unification  way to understand the standard model:}}

Here, we comment briefly on how far
Grand unified theories go and where they fall short in providing
an answer to the same questions discussed above.
 Grand unified theories are attractive for many reasons[17]; among the
reasons for advocating the grand unified theories is one that says that
it provides an answer to the long standing question of charge quantization.
One might also think that these theories may also provide an understanding
of the standard model since they generate at low energies the standard
model as a consequence of symmetry breaking. I argue in this section that
both of these beliefs are not true. To see the point, let us remind
ourselves that
in constructing any typical  GUT theory one makes a series of assumptions.
We find that it is these assumptions
which stand in the way of providing any fundamental understanding of the above
questions. First, one assumes that the fundamental costituent fermions
(i.e. the quarks and the leptons ) belong to a specific representation of the
gauge group : e.g. for SU(5), it is the anomaly free combination
 $\overline{\bf 5}$+{\bf 10};
 in the case of SU(6), it is 2  $\overline{\bf 6}$+{\bf 15} etc.
The question then is what is reason behind this choice. Or to put it
another way, many alternative theories could be constructed by choosing
different anomaly free combinations. Secondly, even given the first
assumption, one  has to additionally
assume specific Higgs representation for
symmetry breaking in order to get the
correct charges for the quarks and leptons
and very easily alternative set of Higgs multiplets could be chosen so
as to yield a different charge assignment for the fermions. For example,if
SU(6) were the GUT group with the usual  $\overline{\bf 6}$+{\bf 15} assignment
of fermions breaking the GUT symmetry by {\bf 6} -Higgs gives the correct
charge assinment of fermions whereas breaking it by {\bf 15} does not. Thus
no real understanding of the electric charges emerges from the grand
unified theories. For some other examples, see ref.19.
 It is worth pointing out that SU(5) and SO(10) appear
to have some very interesting properties in this regard i.e. in the SU(5),
always the correct charge assinment emerges and
 for SO(10) also, I have checked that
Higgs multiplets upto {\bf 560} give the correct assignment of charges once
one breaks the GUT group down to the standard model. In this sense the
string theories provide a very welcome relief in the sense that they not
only predetermine the fermion assignment but also the Higgs multiplets
for symmetry breaking . The problem there is of course the well known
one of non-uniqueness of the gauge group etc due to non-uniqueness of the
vacuum.

\bb

{\bf VI. Conclusion:}

In conclusion, a possible way to derive the standard model for
one generation of fermions from a more economical set of assumptions
than is commonly used, is presented and  it is shown, how in this framework
one can understand such mysteries of nature as the quantization of
 electric charge without necessarily invoking  magnetic monopoles or
the idea of grand unification. In this derivation, the new
concept of an irreducible $Y$-set has been used.
A single family standard model is nothing but a fifteen entry
irreducible $Y$-set.  This concept of irreducible
$Y$-set is quite interesting and may play a role in understanding
the inter-relation between generations.
Applying the same considerations in the presence of a non-vanishing
neutrino mass, it is shown  how the possible Majorana nature of the neutrino
may have a deeper physical meaning conncted with electric charge
quantization.

I like to thank P. Frampton for discussions and collaboration.
This work was supported by a grant from the National Science Foundation.

\bb
\filbreak

\bb

\ce{\bf References}
\b
\item{[1]}
R. E. Marshak and R. N. Mohapatra, {\it Phys. Lett.}{\bf 91B}, 222 (1980).
\item{[2]}
A. Gamba, R. E. Marshak and S. Okubo, {\it Proc. Nat. Acad. Sci. U. S.}
{\bf 45}, 881 (1959).
\item{[3]}
J. D. Bjorken and S. L. Glashow, {\it Phys. Lett.} {\bf 11}, 255 (1964).
\item{[4]}
J. C. Pati and A. Salam, {\it Phys. Rev.} {\bf D10}, 275 (1974);
R. N. Mohapatra and J. C. Pati, {\it Phys. Rev.} {\bf D11}, 566 and
 2558 (1975);
G. Senjanovi\'c and R. N. Mohapatra, {\it Phys. Rev.} {\bf D12}, 1502 (1975).
\item{[5]}
R. N. Mohapatra and G. Senjanovi\'c, {\it Phys. Rev. Lett.} {\bf 44}, 912
(1980).
\item{[6]}
R. N. Mohapatra and R. E. Marshak, {\it Phys. Rev. Lett.} {\bf 44}, 1316
(1980).
\item{[7]}
M. Baldoceolin et al, {\it Phys. Lett.} {\bf 236B}, 95 (1990);
G. Fidecaro et al, {\it Phys. Lett.} {\bf 156B}, 122 (1985);
For a review, see D. Dubbers, {\it Prog. in Part. and Nucl. Physics}
{\it 26}, 173 (1991).
\item{[8]}
R. E. Marshak, {\it Conceptual Foundations of Maodern Particle Physics},
World Scientific, Singapore (1993).
\item{[9]}
S. L. Glashow, {\it Nucl. Phys.} {\bf 22}, 579 (1961).
S. Weinberg, {\it Phys. Rev. Lett.} {\bf 19}, 1264 (1967);
A. Salam, in {\it Elementary Particle Theory } ( edited by N. Swartholm)
Almquist and Forlag, Stockholm, (1968).
\item{[10]} C. Q. Geng and R. E. Marshak, {\it Phys. Rev.}
{\bf D39}, 693 (1989).
\item{[11]}S. L. Adler, {\it Phys. Rev.} {\bf 177}, 2426 (1969);
J. S. Bell and R. Jackiw, {\it Nuov. Cim.} {\bf 51A}, 47 (1969);
W. Bardeen, {\it Phys. Rev.} {\bf 184}, 1848 (1969).
\item{[12]}
 A. Font, L. Ibanez and F. Quevedo, {\it Phys.
Lett.} {\bf 228B}, 79 (1989);
J.Minahan, P. Ramond and R. Warner, {\it Phys. Rev. } {\bf D41}, 715 (1990).
\item{[13]}
 N. G. Deshpande, Oregon Preprint OITS-107 (1979);
R. Foot, G. C. Joshi, H. Lew and R. R. Volkas, {\it Mod. Phys. Lett.}
{\bf A5}, 95 (1990);
\item{[14]}
K.S. Babu and R. N. Mohapatra, {\it Phys. Rev. Lett.} {\bf 63}, 938 (1989);
\item{[15]}
S. Rudaz, {\it Phys. Rev.} {\bf D41}, 2619 (1990).
E. Golowich and P. B. Pal, {\it Phys. Rev.} {\bf D41}, 3537 (1990)
\item{[16]}
D. Gross, J. Harvey, E. Martinec and R. Rohm, {\it Nucl. Phys.} {\bf B256},
253 (1985).
\item{[17]}
P. Frampton and R. N. Mohapatra, UMD-PP-94-72, November, (1993).
\item{[18]}
For a review, see R. Foot,
 G. Joshi, H. Lew and R. Volkas, {\it Mod. Phys. Lett.} {\bf
A5}, 2721 (1990)
\item{[19]}
L. Okun, M. Voloshin and V. I. Zakharov, {\it Phys. Lett.} {\bf 138B},
115 (1984);
R. N. Mohapatra,{\it From Symmetries to Strings: Forty Years of Rochester
Conferences}, ed. A. Das ( World Scientific,1990); p.57.
\bye